\documentclass[10pt,a4paper]{article}
\usepackage{JISE}
\usepackage{graphicx}
\usepackage{times}
\usepackage{mathptmx}

\usepackage{enumitem}
\usepackage{cleveref}
\usepackage{booktabs}
\usepackage{multirow}
\usepackage{amsmath}
\usepackage{amssymb}
\usepackage{url} 

\usepackage{xcolor}
\newif\ifshowchanges
\showchangesfalse
\ifshowchanges
  \newcommand{\add}[1]{\textcolor{red}{#1}}
\else
  \newcommand{\add}[1]{#1}
\fi

\crefname{figure}{Fig.}{Figure}

\title{Generalized Stock Price Prediction for Multiple Stocks Combined with News Fusion}

\author{\normalsize
    Pei-Jun Liao{\small$^{1,2}$}, Hung-Shin Lee{\small$^{5,+}$}, Yao-Fei Cheng{\small$^{4}$}, \\
    \normalsize
    Li-Wei Chen{\small$^{3,5}$}, Hung-yi Lee{\small$^{1}$}, and Hsin-Min Wang{\small$^{2,+}$}\\
  {\small\em$^{1}$Graduate Institute of Electrical Engineering, National Taiwan University, Taipei, Taiwan} \\
  {\small\em$^{2}$Institute of Information Science, Academia Sinica, Taipei, Taiwan} \\
  {\small\em $^{3}$United Link Co., Ltd., Taipei, Taiwan} \\
  {\small\em $^{4}$University of Washington, Seattle, WA, USA} \\
  {\small\em $^{5}$National Tsing Hua University, Hsinchu, Taiwan} \\
}

\authorrunning{Author One, Author Two, Author Three}
\titlerunning{Example for Using the JISE Template}

\begin{document}

\maketitle

\begin{abstract}
\label{sec:abs}
Predicting stock prices presents challenges in financial forecasting.
While traditional approaches such as ARIMA and RNNs are prevalent, recent developments in Large Language Models (LLMs) offer alternative methodologies.
This paper introduces an approach that integrates LLMs with daily financial news for stock price prediction.
To address the challenge of processing news data and identifying relevant content, we utilize stock name embeddings within attention mechanisms.
Specifically, we encode news articles using a pre-trained LLM and implement three attention-based pooling techniques---self-attentive, cross-attentive, and position-aware self-attentive pooling---to filter news based on stock relevance.
The filtered news embeddings, combined with historical stock prices, serve as inputs to the prediction model.
Unlike prior studies that focus on individual stocks, our method trains a single generalized model applicable across multiple stocks.
Experimental results demonstrate a 7.11\% reduction in Mean Absolute Error (MAE) compared to the baseline, indicating the utility of stock name embeddings for news filtering and price forecasting within a generalized framework.
\end{abstract}



\editorfootnote{\small$^+$ Corresponding authors: Hung-Shin Lee and Hsin-Min Wang.}

\section{Introduction}
\label{sec:intro}
Predicting stock prices continues to be a challenge in financial forecasting, remaining an active area of research \cite{chen_2015, nayak_2016, li_2023, wang_2023, bl_2023, kao_2024}.
Recent studies have increasingly integrated textual data with historical stock prices in their predictive models.
Currently, benchmark datasets often use concise social media posts from platforms such as X (formerly Twitter) or Facebook \cite{xu_2018, wu_2018}, utilizing information retrieval techniques to filter and extract relevant content for stock price forecasting.

However, articles obtained through information retrieval techniques often yield a limited quantity of data and may miss implicitly relevant information not easily captured by conventional keyword-based queries, such as those relying solely on stock names.
Furthermore, separating the retrieval phase from the stock price prediction model introduces a dependency on the accuracy and reliability of the independent retrieval process.

Conversely, bypassing retrieval and aggregating all available daily articles introduces significant noise.
This noise complicates processing and analysis, hindering the integration of textual information with stock price data.
Consequently, irrelevant textual information can obscure the prediction process and degrade model accuracy.

To address this problem, we employ an attentive pooling mechanism \cite{vaswani_2017} to filter heterogeneous information.
This mechanism encodes the extracted text into a fixed-length embedding using a pre-trained language model (PLM).
Rather than filtering news for individual stocks a priori \cite{wang_2023}, we aggregate all financial news articles published daily, processing a significantly larger volume of text compared to short social media posts.
Building upon models like BELT \cite{dong_2020} and BigData22 baselines \cite{wu_2018}, which primarily rely on simple concatenation and cross-attention \cite{seo_2016} mechanisms, we systematically evaluate various text and price fusion strategies.
Specifically, we investigate fusion techniques that dynamically weight textual information and market price signals to capture market dynamics.
This integration allows the model to utilize both structured financial signals and unstructured textual data for prediction.

In this work, we demonstrate that news, when combined with the pooling mechanism, can assist in stock price prediction. We selected six individual stocks from the Taiwan Stock Exchange (TWSE) for our experiments.
Furthermore, we conduct additional validation on individual stocks from the BigData23 \cite{wang_2023} dataset, which encompasses the U.S. stock market, to evaluate cross-market applicability.
Unlike previous models trained on individual stocks, our objective is to develop a single generalized model across multiple stocks and evaluate its performance on unseen data.
In a generalized model handling multiple stocks, news filtering requires dynamic, stock-specific context. 
To address this, we incorporate individual stock names into the attention mechanism as queries.
We explore three variants of attention mechanisms to integrate stock name information: self-attentive pooling \cite{safari_2020}, cross-attentive pooling \cite{kye_2021}, and position-aware self-attentive pooling \cite{zhang_2022}.
Our experimental results demonstrate that this approach achieves a 7.11\% reduction in Mean Absolute Error (MAE) compared to the baseline.
In the fusion stage, we build upon cross-attention and concatenation by incorporating Graph Convolutional Networks (GCNs) \cite{chen_2018, zhang_2019}.
These networks model the structural dependencies between stocks.
Furthermore, we utilize bidirectional cross-attention to capture interactions between textual and price signals.
This fused representation is subsequently utilized as input for Time-LLM \cite{jin_2024}, upon which our proposed model is grounded.

This work presents the following contributions:

\begin{enumerate}[noitemsep,leftmargin=*]
\item \textbf{General Stock Price Prediction Model (Sec. 3)}: We introduce a \add{``General'' model, defined as a single unified model trained across multiple stocks.
Unlike previous individual-asset models, this generalized framework captures collective market dynamics and demonstrates applicability across the Taiwan and U.S. markets.}

\item \textbf{Semantic-Aware News Selection Mechanism (Sec. 3-2)}: We propose a pooling mechanism guided by \textit{Stock Name Embeddings}.
This allows the model to leverage the semantic knowledge of Large Language Models (LLMs) to automatically filter irrelevant noise from daily news feeds, addressing the information noise problem in financial text processing.

\item \add{\textbf{Integration of News and Price (Sec. 3-3)}: Integration of News and Price (Sec. 3-3): By integrating GCNs with cross-attention, we fuse unstructured textual representations with structured numerical data.
News is first encoded into contextual text representations by a PLM.
The LLM then serves as the backbone to integrate and transform these text representations together with price series. This enables the predictor to exploit information from unstructured text that is not explicitly present in price histories.}

\item \textbf{Empirical Validation (Sec. 4 \& 5)}: We conduct experiments on both the Taiwan Stock Exchange and the U.S. BigData23 dataset.
The results demonstrate reductions in prediction errors compared to established baselines and provide an analysis of name-based versus ticker-based embeddings in different market contexts.
\end{enumerate}
\section{Related Work}
\label{sec:related_works}

\subsection{Stock Price Prediction}
\label{ssec:stock_prediction_models}

We present a stock price prediction methodology that integrates fusion strategies and attention mechanisms to combine price signals with textual information.

Conventional approaches to stock price prediction often formulate the task using logistic regression \cite{ou_1989} or moving average methodologies \cite{lauren_2014}.
Among these, Autoregressive Integrated Moving Average (ARIMA) models \cite{ariyo_2014, wulff_2017} are widely used for modeling time series through autoregressive and moving average components.

The development of deep learning has led to a range of models for stock price prediction.
Recurrent Neural Networks (RNNs) \cite{kamijo_1990, wang_2023_a} and Long Short-Term Memory networks (LSTMs) \cite{lu_2021, agrawal_2013, khare_2017, chen_2015, hu_2021, pramod_2020, chen_2018, mehtab_2020, wang_2024_a, wang_2024_b, billah_2024} have been used to capture temporal dependencies.
Convolutional Neural Networks (CNNs) have also been applied to model short-term market fluctuations by extracting localized patterns \cite{adebiyi_2014, daiya_2020, mehtab_2020, hoseinzade_2019}.

More recently, Transformer-based models, initially developed for natural language processing (NLP), have been adopted for stock price prediction \cite{li_2023}, alongside deep learning architectures based on multilayer perceptrons (MLPs) \cite{fan_2024, bl_2023}.

Time series forecasting has also advanced with Transformer architectures, including PatchTST \cite{nie_2023, wang_2024} and Autoformer \cite{wu_2022}.
Motivated by the success of large language models (LLMs) in NLP, researchers have proposed models for time series forecasting, including TimesFM \cite{das_2024}, FPT \cite{zhou_2023}, AutoTimes \cite{liu_2024}, and Time-LLM \cite{jin_2024}.
These models leverage generative pre-trained transformer (GPT) \cite{radford_2018, radford_2019, brown_2020, ray_2023} and LLaMA \cite{touvron_2023, touvron_2023_1, grattafiori_2024} architectures.

\subsection{Text Information}
\label{ssec:stock_price_prediction_and_text_information}

Prior work has shown that textual information can complement price-based signals for stock price prediction.
For example, the BELT framework \cite{dong_2020} uses BERT-based \cite{kenton_2019} encoding to extract information from X (formerly Twitter) and integrates it with LSTM-based stock price prediction models.
Similarly, our model integrates textual information with price data to forecast future stock prices.

Textual data from social media and news platforms has been increasingly incorporated into stock price prediction \cite{fung_2003, mohan_2019, xu_2018, wu_2018, wang_2023, soun_2022, gupta_2020, attanasio_2019, bl_2023, lauren_2014}.
Advances in NLP, including large language models and pre-trained language models (PLMs), have improved the extraction and representation of information from financial text.
The BERT family of models \cite{kenton_2019, liu_2019} provides contextualized encodings that support more informative representations of text.

Generative large language models have further expanded the use of text in stock price prediction by providing additional mechanisms to model latent relationships and sentiment linked to market dynamics.
Specialized models such as FinLLMs \cite{lee_2025} and FinGPT \cite{yang_2023}, trained on financial text, facilitate incorporating textual information into forecasting pipelines \cite{xie_2023, tong_2024, wu_2023}.
In this study, we primarily use BERT to encode Chinese-language text and, where applicable, DeBERTa \cite{he_2021} to encode the BigData23 \cite{wang_2023} dataset to support fair comparisons.
Building on prior research, we investigate the integration of price and text information to improve predictive accuracy.
Various studies have explored methodologies to incorporate large language models into stock price prediction.
\section{Proposed Method}
\label{sec:proposed_method}

\begin{figure}[!t]
\centering
\includegraphics[width=1\linewidth]{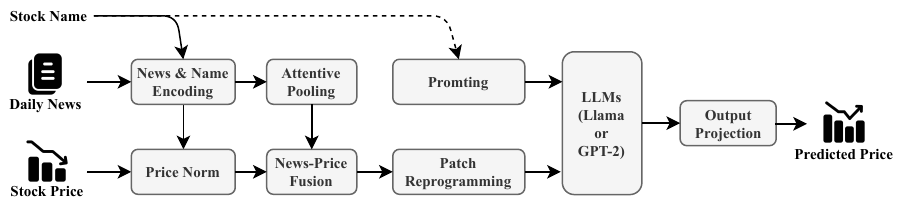}
\vspace{-20pt}
\caption{Overview of the proposed stock price prediction framework.
We employ Attentive Pooling to derive pooled representations of all news articles from each day. Subsequently, we integrate daily historical stock price data with corresponding news information via News-Price Fusion.
Finally, these historical representations are input into an LLM to predict future stock prices. The dotted line represents a process that is optional.}
\label{fig:framework}
\end{figure}

We propose a stock price prediction model that extends the Time-LLM framework \cite{jin_2024} by integrating relevant news data.
Figure~\ref{fig:framework} illustrates the architecture of the proposed model.
For each standard monthly trading window of 20 trading days, we apply separate normalization to (i) daily closing prices and (ii) the news articles associated with each trading day, and then encode the news using a PLM (i.e., News Encoding).
Next, the encoded news is aggregated together with stock name embeddings via a pooling mechanism (i.e., Attentive Pooling) to obtain a daily news representation.
Across the 20 trading days, the resulting sequence of pooled news vectors is fused with the corresponding price sequence using a fusion module (i.e., News-Price Fusion), thereby combining information from both modalities.
The fused sequence representation is then provided to an LLM to predict future stock prices; in this study, we focus on one-day-ahead forecasting.
The following subsections describe the design and functionality of each module.

\subsection{News Encoding}
Each news article is independently encoded using a pre-trained BERT\footnote{\url{https://huggingface.co/ckiplab/bert-base-chinese}} or DeBERTa\footnote{\url{https://huggingface.co/microsoft/deberta-v3-base}} model.
We use the $\mathrm{[CLS]}$ token representation from the final layer of BERT or DeBERTa as the document embedding, which provides a fixed-length semantic representation of the article.

\subsection{Prompting in LLM}
\label{ssec:prompting}

Our input contains 20 days of historical closing prices.
Directly adopting the prompting style of the original Time-LLM model may lead to longer prompts, which can reduce the prominence of the numerical price inputs and affect the model's ability to capture temporal patterns.

To mitigate this issue, we use a concise prompt.
Instead of providing detailed data descriptions or statistical summaries, the prompt contains only the target stock name.
This design serves two purposes.
First, it shortens the prompt, so the numerical inputs remain the primary information used for forecasting.
Second, it conditions the LLM on the stock identity, allowing it to align the input sequence with the corresponding stock-specific context.

\subsection{Attentive Pooling for Daily News Collection}
\label{ssec:news_embedding}

Our dataset contains, on average, over 200 news articles per day.
Encoding all article representations directly within the LLM input can be computationally expensive.
Therefore, we apply an attentive pooling mechanism to summarize each day's news collection into a single vector representation.
Given a sequence of BERT-encoded news embeddings ${\{\mathbf{b}_{t,1},\ldots,\mathbf{b}_{t,n}}\}$, expressed as a matrix $\mathbf{B}_t\in \mathbb{R}^{n\times d}$, where $\mathbf{b}_{t,i}\in \mathbb{R}^{d}$ denotes the $d$-dimensional BERT embedding of the $i$-th news article on day $t$, we compute the daily news representation $\mathbf{c}_t\in\mathbb{R}^{d}$ as
\begin{equation}
\label{eq:att_pool_A}
\mathbf{c}_t = \mathrm{Softmax}(\mathbf{w}\mathbf{B}_t^{\intercal})\mathbf{B}_t,
\end{equation}
where $\mathbf{w}\in\mathbb{R}^{d}$ is a trainable parameter.
This formulation corresponds to dot-product attention, where the keys and values share the same representations, and the query is the trainable vector $\mathbf{w}$ \cite{safari_2020}.
The resulting $\mathbf{c}_t$ is then combined with the stock price embeddings and passed to a fusion module to integrate news and price information.

\begin{figure}[!t]
\centering
\includegraphics[width=1\linewidth]{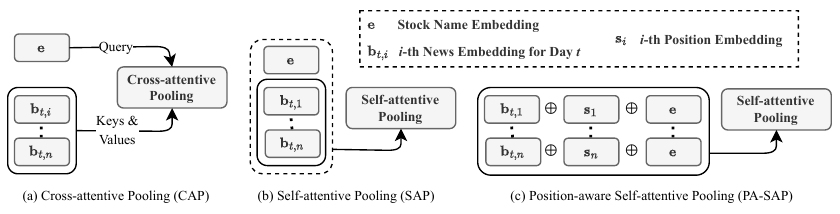}
\vspace{-20pt}
\caption{Three proposed daily news pooling methods that integrate the stock name. (a) We use the sequence of news embeddings as Keys and Values and the stock name embedding as Query for cross-attention to obtain a pooled representation. (b) We append the stock name embedding in front of the sequence of news embeddings, and apply self-attention to obtain a pooled representation. (c) We add the stock name embedding and positional embeddings to the sequence of news embeddings via matrix addition $\oplus$, and apply self-attention to obtain a pooled representation.}
\label{fig:attentive_pooling}
\vspace{-10pt}
\end{figure}

Moreover, the relevance of news articles can differ across stocks.
While prompting can condition the LLM on the target stock, incorporating news embeddings together with stock name embeddings provides an additional mechanism to filter news during inference.
As illustrated in Figure~\ref{fig:attentive_pooling}, we integrate stock name embeddings with features extracted from news articles within the pooling module to produce a stock-conditioned daily news representation.
In contrast to approaches that influence news utilization only at the LLM-based prediction stage, our pooling mechanism incorporates stock name embeddings during the news aggregation step, thereby injecting stock-specific information into the filtering process.

\subsubsection{\textbf{Cross-attentive Pooling (CAP)}}
\label{sssec:cross_att_pooling}
As illustrated in Figure~\ref{fig:attentive_pooling}(a), we use the BERT-encoded stock name embedding, denoted as $\mathbf{e}\in \mathbb{R}^{d}$, as the query in a cross-attention mechanism.
This embedding is obtained by averaging the BERT/DeBERTa token embeddings of the stock name at either the character level or the WordPiece level.
For consistency with the attention formulation used in other pooling mechanisms, we revise \cref{eq:att_pool_A} accordingly.
For day $t$, the news embeddings are denoted as $\mathbf{B}_t\in \mathbb{R}^{n\times d}$ and serve as both keys and values.
The resulting daily representation $\mathbf{c}_t\in\mathbb{R}^{d}$ is computed as
\begin{equation}
\label{eq:cross_att}
\mathbf{c}_t = \mathrm{Softmax}(\mathbf{e}\mathbf{W}_c{\mathbf{B}_t^{\intercal}})\mathbf{B}_t,
\end{equation}
where $\mathbf{W}_c\in\mathbb{R}^{d\times d}$ is a trainable parameter.
This formulation uses the stock name embedding to modulate attention weights over the daily news embeddings $\mathbf{b}_{t,i}$ based on their similarity to the target stock, computed via inner products in the shared embedding space, thereby assigning lower weights to less relevant news articles.

\subsubsection{\textbf{Self-attentive Pooling (SAP)}}
\label{sssec:self_att_pooling}

As shown in Figure~\ref{fig:attentive_pooling}(b), we concatenate the stock name embedding with the news embeddings along the sequence dimension.
We then apply self-attentive pooling to obtain a daily aggregate news representation:
\begin{equation}
\label{eq:self_att}
\mathbf{c}_t = \mathrm{Softmax}(\mathbf{w}_s\tilde{\mathbf{B}}_t^{\intercal}){\tilde{\mathbf{B}}_t},
\end{equation}
where $\tilde{\mathbf{B}}_t=[\mathbf{e};\mathbf{B}_t]\in\mathbb{R}^{(n+1)\times d}$ treats the stock name as an additional item in the sequence, and $\mathbf{w}_s\in\mathbb{R}^{d}$ is a trainable parameter.
Under this formulation, the attention weights are determined jointly by the stock name embedding and the news embeddings, reflecting both stock--news relations and interactions among the news articles.

\subsubsection{\textbf{Position-aware Self-attentive Pooling (PA-SAP)}}
\label{sssec:posself_att_pooling}
As depicted in Figure~\ref{fig:attentive_pooling}(c), we incorporate positional embeddings to account for temporal order within each day's news sequence.
Let $\mathbf{S}\in\mathbb{R}^{n \times d}$ denote sinusoidal positional embeddings as in the Transformer architecture.
We replicate the stock name embedding $\mathbf{e}$ to match the length of $\mathbf{B}_t$, yielding $\mathbf{E}\in\mathbb{R}^{n\times d}$.
We then add positional embeddings to both the replicated stock embedding and the news embeddings, and form the modified embeddings $\bar{\mathbf{B}}_t$.
The resulting sequence is summarized using a self-attentive pooling mechanism, referred to as position-aware self-attentive pooling:
\begin{equation}
\label{eq:pos_aware}
\mathbf{c}_t=\mathrm{Softmax}(\mathbf{w}_p\bar{\mathbf{B}}_t^{\intercal})\bar{\mathbf{B}}_t,
\end{equation}
where $\bar{\mathbf{B}}_t=\mathbf{B}_t+\mathbf{E}+\mathbf{S}$, and $\mathbf{w}_p\in\mathbb{R}^{d}$ is a trainable parameter.

\subsection{News-Price Fusion}
\label{ssec:stock_aware_features_fusion}

\begin{figure}[!t]
\centering
\includegraphics[width=0.5\linewidth]{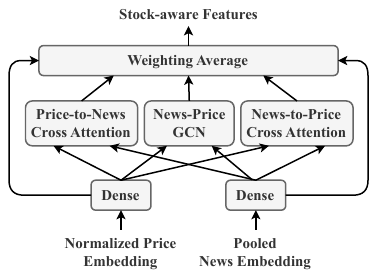}
\vspace{-10pt}
\caption{The News-Price Fusion Module. Initially, we feed the 20-day price embeddings and pooled news embeddings into distinct dense layers.
Subsequently, we apply bidirectional cross-attention, alternating between prices and news as Query/Key and Value.
Additionally, a two-layer GCN architecture is employed to model the interaction between prices and news.
Finally, we integrate the three fused representations with the original outputs from the dense layers (both price and news embeddings) using a weighted average, yielding the final fused result.}
\label{fig:fusion}
\end{figure}

To integrate textual and price information, we use three fusion methods to combine data from news articles and stock prices, generating Stock-aware Features, as illustrated in \Cref{fig:fusion}.

\subsubsection{\textbf{Cross-attention Fusion}}
\label{sssec:cross_att_fusion}

We use cross-attention to integrate daily news embeddings and stock price embeddings.
Specifically, we apply both price-to-news and news-to-price cross-attention \cite{seo_2016}.
Cross-attention is defined as
\begin{equation}
\label{eq:cross_attention}
\mathrm{CrossAtt}(\mathbf{Q}, \mathbf{K}, \mathbf{V}) =
\mathrm{Softmax} \left( \frac{\mathbf{Q} \mathbf{K}^\intercal}{\sqrt{d}} \right) \mathbf{V},
\end{equation}
where $\mathbf{Q} \in \mathbb{R}^{T_Q \times d}$ and $\mathbf{K}, \mathbf{V} \in \mathbb{R}^{T_K \times d}$.
Here, $T_Q$ denotes the length of the query sequence and $T_K$ denotes the length of the key (and value) sequence; the key and value sequences share the same length $T_K$.
Before applying attention, $\mathbf{Q}$, $\mathbf{K}$, and $\mathbf{V}$ are each projected by separate linear layers.

Accordingly, the price-to-news ($\mathbf{S_{P2N}}$) and news-to-price ($\mathbf{S_{N2P}}$) representations are computed as
\begin{equation}
\label{eq:cross_P2N}
\mathbf{S_{P2N}}=\mathrm{CrossAtt}(\mathbf{P}, \mathbf{N}, \mathbf{N}),\;\mathbf{S_{N2P}}=\mathrm{CrossAtt}(\mathbf{N}, \mathbf{P}, \mathbf{P}),
\end{equation}
where $\mathbf{N} = (\mathbf{c}_t \in \mathbb{R}^{d} \mid t = 1, \dots, T)$ and $\mathbf{P} = (\mathbf{p}_t \in \mathbb{R}^{d} \mid t = 1, \dots, T)$ denote the sequences of news embeddings and price embeddings over $T=20$ days (a monthly trading period), respectively.

$\mathbf{S_{P2N}}$ and $\mathbf{S_{N2P}}$ are the outputs of cross-attention under two query--key--value configurations.
Specifically, $\mathbf{S_{P2N}}$ is obtained by using $\mathbf{P}$ as the query and $\mathbf{N}$ as both the key and the value.
Conversely, $\mathbf{S_{N2P}}$ is obtained by using $\mathbf{N}$ as the query and $\mathbf{P}$ as both the key and the value.

Both $\mathbf{S_{P2N}}$ and $\mathbf{S_{N2P}}$ are matrices in $\mathbb{R}^{T \times d}$.
Before applying attention, $\mathbf{N}$ and $\mathbf{P}$ are each passed through separate linear transformation layers to produce their corresponding query, key, and value representations.
These two directions of cross-attention capture dependencies between news and stock prices and allow information to be exchanged between the two modalities.

\subsubsection{\textbf{GCN Fusion}}
\label{sssec:gcn_fusion}

In addition to cross-attention, we incorporate a two-layer GCN \cite{chen_2018, zhang_2019} to integrate news and price information.
The first layer fuses the daily news embeddings and price embeddings to model within-day relations.
The second layer applies a $\mathrm{CausalCNN}$ (Causal Convolutional Neural Network) to aggregate the resulting news-aware price features over a 5-day window.
We use a 5-day window to match the horizon of a five-day short-term moving average, which provides a standard time scale for summarizing short-term price movements.
The formulation is
\begin{equation}
\label{eq:GCN_l1}
\mathbf{H}=\mathrm{GCNConv}(\mathbf{N}, \mathbf{P}),\;\mathbf{S_{GCN}} = \mathrm{CausalCNN}\left( \mathbf{H} \right).
\end{equation}
The input to the first layer consists of the news collection embeddings $\mathbf{N}$ and normalized price embeddings $\mathbf{P}$ over $T$ days, as introduced in \cref{sssec:cross_att_fusion}.
The $\mathrm{GCNConv}$ layer aggregates information within this $T$-day window by connecting the news and price features.
In the second layer, we apply a $\mathrm{CausalCNN}$ with kernel size 5 to the output of the first layer, $\mathbf{H}$.
The causal convolution restricts the receptive field so that, at each time step, the model depends only on past and current inputs, thereby preserving temporal causality.
This convolution aggregates information from five consecutive days to represent short-term temporal dynamics.
The output of this layer is denoted as $\mathbf{S_{GCN}}$.

Finally, we compute a learnable weighted sum of the representations from the fusion methods, together with the original news collection and price embeddings, to obtain a unified representation (Stock-aware Features), as illustrated at the top of \Cref{fig:fusion}.
This representation is then fed into the patch-reprogramming module, where it is concatenated with the prompt within the input sequence for future price prediction.

\subsection{Patch Reprogramming and LLM}
\label{ssec:patch_reprogramming_and_llm} 

LLMs are pre-trained on large-scale text corpora and can model patterns and relationships in natural language.
When applying LLMs to quantitative domains such as finance, a key difficulty is the modality gap between numerical time-series inputs and the discrete, token-based representations used by language models.
To mitigate this gap, we adopt the patch reprogramming technique of Time-LLM \cite{jin_2024}, which enables a frozen LLM to process stock-related features without modifying its core architecture.

The main idea is to map stock-aware features into a representation that is compatible with the LLM embedding space.
We first construct a compressed representation of the LLM vocabulary embedding space.
Specifically, a linear layer projects the vocabulary embeddings from the original size $V$ to a lower dimension $U$ (with $V \gg U$), producing a set of \textit{prototypes} that summarize the vocabulary embedding space.

Given these prototypes, we use cross-attention to map stock-aware features into the LLM embedding space.
In this mechanism, the stock-aware features serve as the \textit{query}, while the vocabulary prototypes serve as both the \textit{key} and the \textit{value}.
This mapping yields a prototype-weighted representation of the input features in the LLM embedding space.
With patch reprogramming, stock-related signals (e.g., price dynamics and news-derived features) are represented in a token-compatible embedding form for subsequent processing by the frozen LLM.
To preserve the pre-trained parameters, we keep the LLM frozen and fine-tune only the reprogramming layers and the final prediction layer.
This design limits the number of trainable parameters during adaptation to the forecasting task.
\section{Experimental Setup}
\label{sec:experiments}

\subsection{Datasets}

\begin{table}[!t]
\centering
\setlength{\tabcolsep}{22pt}
\caption{Maximum and average number of characters in news articles and maximum and average number of news articles per day.
For excessively long articles, methods such as sliding window or truncation are applied.}
\label{tab:data_info}
\vspace{5pt}
\begin{tabular}{cccc}
\toprule
Max \# Chars & Avg \# Chars & Max \# News & Avg \# News  \\ \midrule
7538 & 509.96 & 654 & 219.50 \\ 
\bottomrule
\end{tabular}
\vspace{0pt}
\end{table}

\subsubsection{Taiwan Stock Market Data Collection}
\label{ssec:tw_data}

We collected Taiwan stock market data (TW21) from July~1,~2021 to April~30,~2024 using the twstock API\footnote{\url{https://twstock.readthedocs.io/zh-tw/latest/}}, which contains 1,018 trading days.
We conducted experiments on six stocks: TSMC, MediaTek, Foxconn, Realtek, Novatek, and Delta.
Following the time series partitioning strategy used in the Time-Series Library\footnote{\url{https://github.com/thuml/Time-Series-Library}} and Time-LLM, we split the dataset into training (70\%), validation (10\%), and test (20\%) sets, corresponding to 712, 103, and 203 trading days, respectively.
We also collected Taiwanese financial news from the same period and filtered articles related to finance, consumer markets, and technology.
On average, the dataset contains 219.50 articles per day, and each article has 569.96 Chinese characters (the PLM encodes at the character level), as summarized in \Cref{tab:data_info}.
The prediction task uses 20 days of historical prices and the corresponding news as input (approximately one trading month) and predicts the closing price of the next trading day.

\subsubsection{BigData23 U.S. Stock Market Data}
As an additional analysis across markets, we conducted experiments on 42 U.S. stocks using the BigData23 \cite{wang_2023} dataset\add{, which includes companies from multiple sectors such as technology, finance, and consumer goods.
This composition allows us to examine whether the model relies on broad market signals rather than sector-specific patterns.}
Following the same time series partitioning strategy, we split the data into training (70\%), validation (10\%), and test (20\%) sets.
For each stock, the Twitter texts within a single day were aggregated, and their chronological order was removed.
The resulting daily texts were then aligned with each stock's time sequence, while texts from different stocks were sorted alphabetically and merged into a single day's news input.

\subsection{Model and Training Configurations}
\label{ssec:model_and_environment}

To develop a unified model that generalizes across multiple stocks, we aggregate data from all six stocks in the Taiwan dataset for training and validation, and evaluate the model on the individual test set of each stock.
Similarly, for the BigData23 U.S. market dataset, we combine data from all 42 stocks and align Twitter comments in chronological order to preserve the temporal correspondence between stock prices and social media discussions.
\add{The model was trained to minimize the Mean Squared Error (MSE) loss function.}

In our experiments on the Taiwan dataset, the backbone LLM is the LLaMA3-8b model from the TAIDE project (Llama3-TAIDE-LX-8B-Chat-Alpha1, 8B parameters)\footnote{\url{https://huggingface.co/taide/Llama3-TAIDE-LX-8B-Chat-Alpha1}}.
To examine the dependence on the backbone, we also run experiments with CKIP's GPT-2 (gpt2-base-chinese, 102M parameters)\footnote{\url{https://huggingface.co/ckiplab/gpt2-base-chinese}} as an alternative LLM.
For the BigData23 dataset, we follow the LLaMA and GPT-2 configurations used in Time-LLM, where the backbone LLM is kept frozen during training.

We use early stopping with a patience of $5$ and set the maximum number of epochs to $15$.
The batch size is $4$, and the learning rate is $0.01$ in all experiments.
Other settings follow Time-LLM.
All experiments are conducted on an NVIDIA RTX-3090 GPU with 24GB memory.

\subsection{Evaluation Metrics}
\label{ssec:metrics}
We use Mean Absolute Error (MAE) and Mean Squared Error (MSE) as evaluation metrics by comparing model predictions with the corresponding ground-truth closing prices.
In all experiments, stock prices are normalized using Standard Scaling\footnote{\url{https://scikit-learn.org/dev/modules/generated/sklearn.preprocessing.StandardScaler.html}} during training and inference, and MAE and MSE are computed on the normalized prices.
\section{Experimental Results}
\label{sec:results}

\subsection{Prediction on TW21}
\label{ssec:prediction_stocks_tw}

\begin{table}[t!]
\centering
\setlength{\tabcolsep}{5.4pt}
\caption{Performance of individual stocks in future-1 day prediction. TimeLLM represents the Time-LLM baseline without news, +News represents including news but excluding the stock name, +CAP, +SAP, and +PA-SAP are the proposed news pooling methods integrating the stock name: cross-attentive pooling, self-attentive pooling, and position-aware self-attentive pooling, respectively. \textbf{Bold} text indicates that it surpasses both TimeLLM and +News, and underline indicates the best in each column.}
\label{tab:genera_future1}
\vspace{5pt}
\begin{tabular}{l|c|cc|cc|ccc}
\toprule
\multirow{2}{*}{Company} & \multirow{2}{*}{Metric} & \multicolumn{7}{c}{Methods} \\
\cmidrule(lr){3-9}
& & LSTM & FPT & TimeLLM & +News & +CAP & +SAP & +PA-SAP \\ 
\midrule
\multirow{2}{*}{TSMC} 
  & MAE & 0.2677 & 0.1477 & 0.1496 & 0.1530 & \underline{\textbf{0.1398}} & \textbf{0.1433} & \textbf{0.1419} \\
  & MSE & 0.2000 & 0.0457 & 0.0456 & 0.0446 & \underline{\textbf{0.0399}} & \textbf{0.0429} & \textbf{0.0432} \\ 
\midrule
\multirow{2}{*}{MediaTek} 
  & MAE & 0.2931 & 0.1403 & 0.1365 & 0.1347 & \textbf{0.1276} & \textbf{0.1294} & \underline{\textbf{0.1200}} \\ 
  & MSE & 0.2270 & 0.0356 & 0.0340 & 0.0346 & \textbf{0.0292} & \textbf{0.0301} & \underline{\textbf{0.0261}} \\ 
\midrule
\multirow{2}{*}{Foxconn} 
  & MAE & 1.0633 & 0.5378 & 0.5016 & 0.5035 & \underline{\textbf{0.4763}} & \textbf{0.4953} & 0.5125 \\ 
  & MSE & 0.8225 & 0.7734 & 0.6935 & 0.7776 & \underline{\textbf{0.6479}} & 0.7216 & 0.7237 \\ 
\midrule
\multirow{2}{*}{Realtek} 
  & MAE & 0.2449 & 0.0930 & 0.0892 & 0.0961 & 0.0921 & \underline{\textbf{0.0880}} & 0.0898 \\ 
  & MSE & 0.1752 & 0.0175 & 0.0164 & 0.0192 & 0.0167 & \textbf{0.0161} & \underline{\textbf{0.0147}} \\ 
\midrule
\multirow{2}{*}{Novatek} 
  & MAE & 0.2727 & \underline{0.0890} & 0.0938 & 0.0949 & \textbf{0.0929} & \textbf{0.0921} & \textbf{0.0910} \\ 
  & MSE & 0.2152 & 0.0167 & 0.0171 & 0.0189 & \textbf{0.0169} & \textbf{0.0159} & \underline{\textbf{0.0157}} \\ 
\midrule
\multirow{2}{*}{Delta} 
  & MAE & 0.3536 & 0.1938 & 0.1887 & 0.1781 & \underline{\textbf{0.1740}} & \underline{\textbf{0.1740}} & \textbf{0.1772} \\ 
  & MSE & 0.2614 & 0.0763 & 0.0673 & 0.0613 & \textbf{0.0612} & \underline{\textbf{0.0564}} & \textbf{0.0581} \\ 
\bottomrule
\end{tabular}
\vspace{-10pt}
\end{table}

The first experiment evaluates a future-1 day prediction scenario, where a model trained on a mixed-stock dataset is tested across six individual stocks.
Two Time-LLM baselines are considered: (1) without news (TimeLLM) and (2) with news but without stock name embeddings (+News).
Additionally, we compare against two classical models: a pure LSTM and an FPT model.
As demonstrated in \Cref{tab:genera_future1}, cross-attentive pooling (+CAP) and self-attentive pooling (+SAP) yield lower MAE and MSE compared to the baselines across the majority of individual stocks.

In the TW21 dataset, we observe that among all pooling methods, +CAP yields the lowest MSE for TSMC (0.0399) and Foxconn (0.6479), and the lowest MAE for both TSMC (0.1398) and Foxconn (0.4763).

+SAP outperforms the baselines across most cases, though it yields the lowest overall error on fewer individual stocks than +CAP.
These results indicate that, within TW21, +CAP minimizes errors for large-cap stocks such as TSMC and Foxconn, whereas +SAP exhibits lower error variance across the different individual stocks.

Position-aware self-attentive pooling (+PA-SAP) yields the lowest errors for specific stocks, particularly MediaTek, where it attains the lowest MAE (0.1200) and MSE (0.0261), and Novatek, where it achieves the lowest MSE (0.0157).
However, its error metrics show higher variance across the dataset compared to +SAP.

Overall, these findings indicate that within the TW21 dataset, the proposed pooling mechanisms reduce prediction errors across the evaluated stocks.
\add{In addition, multi-seed evaluation (\Cref{tab:seed_combined}) shows that while +CAP yields the lowest mean errors in large-cap stocks, +SAP yields comparable mean errors with lower or equal standard deviations across most individual stocks.
The relative ranking between +CAP and +SAP remains unchanged across seeds, indicating that the observed error reductions are not driven by random initialization.
These results suggest that +SAP exhibits lower sensitivity to initialization variance.}

\begin{table}[t]
\centering
\scriptsize
\setlength{\tabcolsep}{5.4pt}
\caption{
\add{Mean performance over \textbf{three independent random seeds} with \textbf{sample} standard deviation in parentheses.
The upper block reports MAE and the lower block reports MSE. Lower is better.}
}
\label{tab:seed_combined}
\begin{tabular}{lcccccc}
\toprule
~ & TSMC & MediaTeK & Foxconn & Realtek & Novatek & Delta \\
\midrule
\multicolumn{7}{c}{MAE Mean (Std)} \\
\midrule
TimeLLM    & 0.1502 (0.0099) & 0.1335 (0.0104) & 0.5092 (0.0474) & 0.0909 (0.0047) & 0.0931 (0.0011) & 0.1876 (0.0080) \\
+News    & 0.1444 (0.0077) & 0.1295 (0.0121) & 0.5153 (0.0409) & 0.0924 (0.0061) & 0.0923 (0.0023) & 0.1815 (0.0044) \\
+SAP    & 0.1414 (0.0043) & 0.1227 (0.0078) & 0.4927 (0.0138) & 0.0853 (0.0023) & 0.0894 (0.0024) & 0.1744 (0.0099) \\
+CAP   & 0.1390 (0.0007) & 0.1252 (0.0090) & 0.5019 (0.0408) & 0.0897 (0.0040) & 0.0933 (0.0014) & 0.1807 (0.0074) \\
+PA-SAP & 0.1489 (0.0145) & 0.1333 (0.0116) & 0.5510 (0.0345) & 0.0967 (0.0068) & 0.0936 (0.0049) & 0.1889 (0.0139) \\
\midrule
\multicolumn{7}{c}{MSE Mean (Std)} \\
\midrule
TimeLLM    & 0.0463 (0.0048) & 0.0327 (0.0044) & 0.7285 (0.1439) & 0.0166 (0.0018) & 0.0174 (0.0004) & 0.0670 (0.0070) \\
+News    & 0.0441 (0.0009) & 0.0305 (0.0061) & 0.7659 (0.1430) & 0.0174 (0.0022) & 0.0177 (0.0014) & 0.0600 (0.0013) \\
+SAP    & 0.0432 (0.0024) & 0.0271 (0.0034) & 0.6902 (0.0571) & 0.0150 (0.0010) & 0.0164 (0.0004) & 0.0562 (0.0045) \\
+CAP   & 0.0418 (0.0017) & 0.0282 (0.0037) & 0.7264 (0.1399) & 0.0163 (0.0010) & 0.0173 (0.0007) & 0.0613 (0.0010) \\
+PA-SAP & 0.0482 (0.0067) & 0.0321 (0.0053) & 0.9120 (0.1888) & 0.0185 (0.0039) & 0.0174 (0.0026) & 0.0650 (0.0108) \\
\bottomrule
\end{tabular}
\end{table}

\begin{table}[t!]
\centering
\setlength{\tabcolsep}{22pt}
\caption{Performance on TW21 for future-1 day prediction. TimeLLM represents the baseline without news, while +News includes news but excludes the stock name. +CAP, +SAP, and +PA-SAP are proposed news pooling methods integrating the stock name. \textbf{Bold} text denotes surpassing both TimeLLM and +News, and underline highlights the best in each column.}
\label{tab:avg_analysis}
\vspace{10pt}
    \begin{tabular}{l|llll}
        \toprule
        \multirow{2}[2]{*}{Method} & \multicolumn{2}{c}{LLaMA Backbone} & \multicolumn{2}{c}{GPT-2 Backbone} \\ 
        \cmidrule(lr){2-3} \cmidrule(lr){4-5}
        & MAE & MSE & MAE & MSE \\ 
        \midrule 
        \multicolumn{5}{c}{\textbf{without SNP}} \\
        \midrule  
        TimeLLM & 0.1913 & 0.1497 & 0.1889 & 0.1380  \\ 
        +News & 0.1874 & 0.1432 & 0.1849 & 0.1391  \\ 
        \cmidrule[\lightrulewidth](lr){1-5} 
        +CAP & \textbf{0.1830} & \underline{\textbf{0.1252}} & \textbf{0.1775} & \textbf{0.1200}  \\ 
        +SAP & \underline{\textbf{0.1777}} & \textbf{0.1311} & \textbf{0.1756} & \textbf{0.1191}  \\ 
        +PA-SAP & 0.1936 & 0.1535 & \textbf{0.1788} & \textbf{0.1321}  \\ 
        \midrule
        \multicolumn{5}{c}{\textbf{with SNP}} \\ 
        \midrule
        TimeLLM & 0.1932 & 0.1457 & 0.2000 & 0.1574  \\ 
        +News & 0.1934 & 0.1594 & 0.1725 & 0.1170  \\ 
        \cmidrule[\lightrulewidth](lr){1-5} 
        +CAP & \textbf{0.1838} & \textbf{0.1353} & \underline{\textbf{0.1704}} & \underline{\textbf{0.1106}}  \\ 
        +SAP & \textbf{0.1870} & 0.1472 & 0.1765 & 0.1276  \\ 
        +PA-SAP & \textbf{0.1887} & 0.1469 & 0.1752 & 0.1256  \\ 
        \bottomrule
    \end{tabular}
\vspace{-10pt}
\end{table}

\subsection{Pooling Methods and Stock Name Prompts Across LLMs}
\label{ssec:pooling_prompt_llms}

To evaluate the various pooling mechanisms, we first analyze the LLaMA backbone by averaging the test MAE and MSE across all stocks.
As shown in \Cref{tab:avg_analysis}, +SAP reduces prediction errors.
Without the Stock Name Prompt (SNP), +SAP reduces MAE from 0.1913 to 0.1777 (7.11\%) and MSE from 0.1497 to 0.1311 (12.42\%), yielding lower errors than both the TimeLLM and +News baselines.

With SNP, the error reductions from +SAP are less pronounced.
MAE decreases from 0.1932 to 0.1870 (3.21\%), while MSE slightly increases from 0.1457 to 0.1472 (1.03\%).
Compared to +News, +SAP reduces MAE and MSE by 3.31\% and 7.65\%, respectively.
In this setting, +CAP yields the lowest MSE (0.1353) with an MAE of 0.1838.
This indicates that the interactions captured by +CAP between stock-related features and news contribute to lower prediction errors.

We further evaluate position-aware self-attentive pooling (+PA-SAP).
Without SNP, +PA-SAP achieves MAE and MSE values of 0.1788 and 0.1321, respectively; while these are lower than the +News baseline, they do not consistently drop below the errors of +SAP or +CAP.
Specifically, although +PA-SAP achieves a lower MAE compared to +CAP (0.1788 vs. 0.1830), it results in a higher MSE (0.1321 vs. 0.1252).
With SNP, +PA-SAP yields an MAE of 0.1752 and an MSE of 0.1256, which remains higher than the MSE of +CAP.

A potential reason for the performance of +PA-SAP is that causal relationships among different news articles may not be distinct enough to benefit from positional encoding.
Unlike structured financial reports or explicitly linked event sequences, financial news articles often exhibit weak or implicit causal dependencies.
Consequently, incorporating positional relationships does not further reduce prediction errors.
One direction for future work is to pre-filter news articles based on causal relationships before applying pooling mechanisms.
Identifying causally linked news items and prioritizing them in the pooling process could assist in capturing market signals.

To verify that these trends are not model-specific, we extended our analysis to CKIP-GPT2.
The results indicate that +SAP reduces MAE/MSE by 7.04\%/13.70\% without SNP.
When SNP is included, +SAP yields MAE/MSE reductions of 11.75\%/18.93\%.
However, under CKIP-GPT2, +SAP yields higher errors than +CAP, which achieves the lowest overall errors.
This suggests that +CAP captures interactions between stock and news information, particularly when using a backbone with lower prompt-following capabilities.

+SAP exhibits lower error variance across settings, particularly without SNP.
It yields consistent error metrics, whereas other methods exhibit higher variance.
Furthermore, under the LLaMA backbone, the absence of SNP leads to larger error reductions, likely because stock name embeddings already provide sufficient contextual information, reducing the requirement for additional disambiguation through prompt-based mechanisms.

\begin{table}[t!]
    \centering
    \setlength{\tabcolsep}{18pt}
    \caption{Ablation study results for the +SAP method. The ``-'' symbol indicates the removal of specific components from the News-Price Fusion module in Figure \ref{fig:framework}. Underlined values represent the best performance in each column.}
    \label{tab:ablation}
    \vspace{5pt}
    \begin{tabular}{l|cccc}
        \toprule
        \multirow{2}[2]{*}{Method} & \multicolumn{2}{c}{with SNP} & \multicolumn{2}{c}{without SNP} \\
        \cmidrule(lr){2-3} \cmidrule(lr){4-5}
        ~ & MAE & MSE & MAE & MSE \\
        \midrule
        +SAP & 0.1870 & 0.1472 & \underline{0.1777} & \underline{0.1311} \\
        - GCN & 0.1835 & 0.1420 & 0.1940 & 0.1583 \\
        - P2N  & 0.1882 & 0.1538 & 0.1934 & 0.1611 \\
        - N2P & 0.2166 & 0.2125 & 0.2112 & 0.1889 \\
        - P2N - N2P & 0.2149 & 0.2191 & 0.2145 & 0.2147 \\
        - N2P - GCN & 0.1902 & 0.1605 & 0.1808 & 0.1351 \\
        - P2N - GCN & \underline{0.1776} & \underline{0.1313} & 0.1808 & 0.1341 \\
        - P2N - N2P - GCN & 0.1865 & 0.1421 & 0.1871 & 0.1444 \\
        \bottomrule
    \end{tabular}
    \vspace{-10pt}
\end{table}

\subsection{Ablation Study}
\label{ssec:ablation}

In the ablation studies, we analyze +SAP under the without SNP setting on the TAIDE-LLaMA3 backbone, as detailed in \Cref{tab:ablation}.
The results indicate that, under this scenario, the integration of all fusion mechanisms yields the lowest prediction errors.
Each component contributes to error reduction; removing the GCN layer alone results in a larger increase in error compared to the simultaneous removal of P2N and GCP.

In contrast, under the SNP setting, the N2P-only configuration yields the lowest errors. This indicates that when prompts are present, the additional fusion mechanisms do not further reduce prediction errors.

\begin{figure}[!t]
\centering
\includegraphics[width=0.9\linewidth]{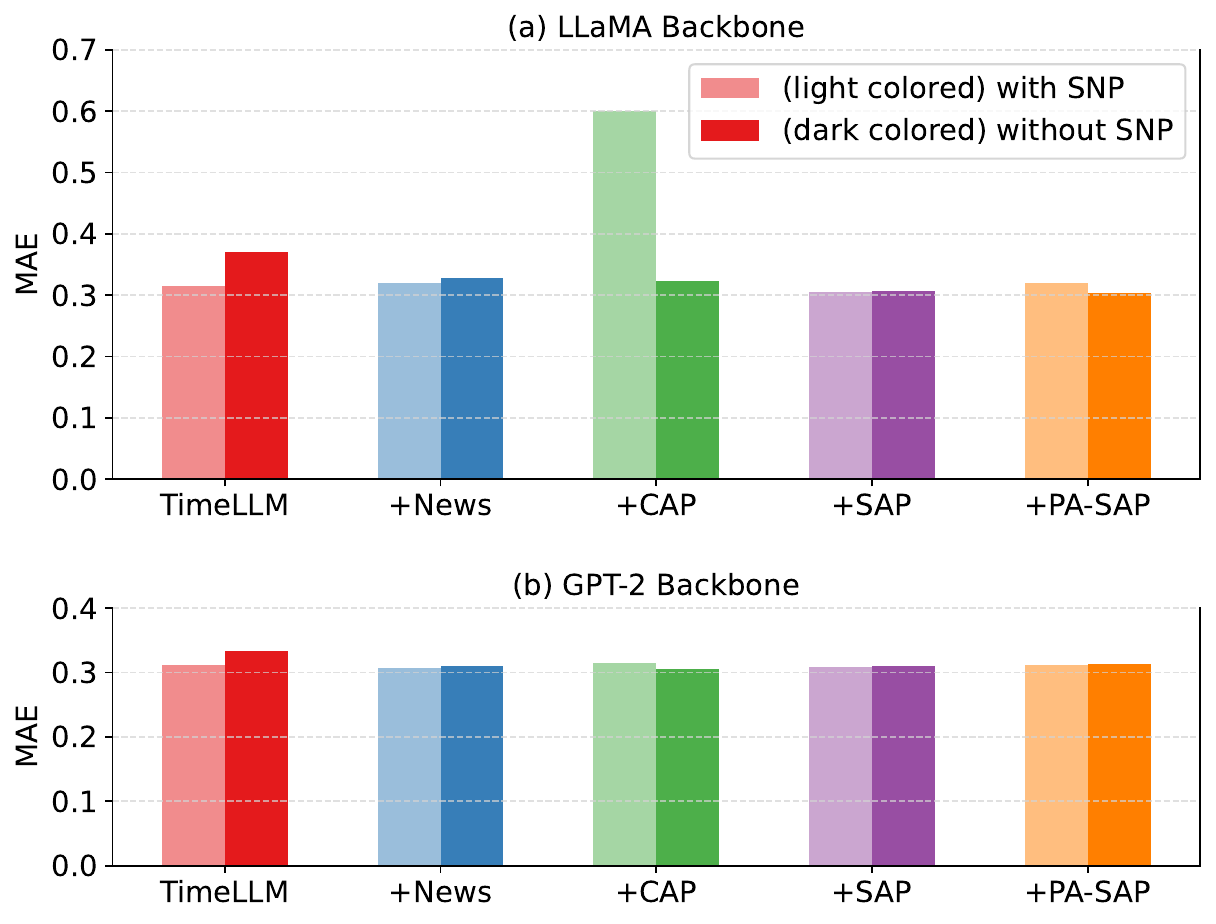}
\vspace{-10pt}
\caption{(a) and (b) compare the MAE of future 5-day price prediction methods with different backbones: LLaMA (a) and GPT-2 (b).  
Darker bars indicate results without SNP, while lighter bars show results with SNP.  
TimeLLM serves as the baseline, +News includes news without stock name embeddings, and +CAP, +SAP, and +PA-SAP are three attentive pooling methods.
}
\label{fig:pred_future}
\vspace{-10pt}
\end{figure}

\subsection{Future-5 Day Prediction}
\label{ssec:future_5_day}

We analyze the average MAE for future-5 day predictions across two backbone models and three attentive pooling mechanisms.
As shown in \Cref{fig:pred_future}, +SAP yields the lowest MAE for short-term forecasting, while +CAP exhibits higher error variance across different stocks.
Incorporating the SNP reduces prediction errors, indicating its utility in leveraging LLMs for sequence modeling over extended horizons.
Exhibiting lower mean errors and lower variance, +SAP provides more consistent forecasting results across the dataset.
These findings suggest that the current 20-day input window aligns with short-term forecasting objectives, whereas extending it to 60 days could be evaluated for long-term trend predictions in future work.

\subsection{BigData23 Results}

\add{The linguistic characteristics of the BigData23 dataset differ from TW21, presenting a challenge for semantic-based models.
While TW21 employs full company names that serve as semantic anchors, BigData23 relies on abbreviated ticker symbols (e.g., ``AAPL'', ``F'').
In the pre-trained embedding space of a PLM, single-letter tickers like ``F'' (Ford) or ``T'' (AT\&T) exhibit polysemy and lack the specific entity associations found in full corporate names.
This semantic ambiguity reduces the precision of name-based entity recognition.}

Following the methodology of Time-LLM \cite{jin_2024}, we evaluated the model using LLaMA2 \cite{touvron_2023_1} and OpenAI's GPT-2 \cite{radford_2019}, focusing on the \textit{without SNP} setting for consistency with TW21.
As shown in \Cref{tab:bigdata23_future1}, +SAP yields the lowest prediction errors under the LLaMA backbone.
Conversely, +CAP exhibits an increase in error metrics.

This divergence provides insight into the attention mechanisms: +CAP relies on the stock name embedding acting as a ``query'' to filter news.
When the ticker embedding is semantically ambiguous, the error reductions from this filtering mechanism diminish.
In contrast, +SAP, which fuses information via self-attention across the concatenated sequence, reduces dependency on the explicit semantic representation of the stock identifier.
These findings suggest that while our framework is applicable across markets, minimizing prediction errors on ticker-based datasets may require mapping abbreviations to full entity names to provide contextual precision.

\begin{table}[t!]
\centering
\setlength{\tabcolsep}{22pt}
\caption{Performance on BigData23 for future-1 day prediction (without SNP). TimeLLM is the baseline without news, while +News includes news but excludes the stock name. +SAP, +CAP, and +PA-SAP are news pooling methods integrating the stock name. \textbf{Bold} surpasses both TimeLLM and +News, and \underline{underline} highlights the best in each column.}
\label{tab:bigdata23_future1}
\vspace{10pt}
\begin{tabular}{l|llll}
    \toprule
    \multirow{2}[2]{*}{Methods} & \multicolumn{2}{c}{LLaMA Backbone} & \multicolumn{2}{c}{GPT-2 Backbone} \\ 
    \cmidrule(lr){2-3} \cmidrule(lr){4-5}
     & MAE & MSE & MAE & MSE  \\ 
    \midrule
    TimeLLM & 0.1030 & 0.0221 & 0.1037 & 0.0222  \\ 
    +News & 0.0975 & 0.0199 & 0.1034 & 0.0223  \\ 
    \cmidrule[\lightrulewidth](lr){1-5} 
    +CAP & 0.2653 & 0.1291 & 0.2443 & 0.1102  \\ 
    +SAP & \underline{\textbf{0.0974}} & \underline{\textbf{0.0198}} & \underline{\textbf{0.1024}} & \underline{\textbf{0.0218}}  \\ 
    +PA-SAP & 0.1019 & 0.0215 & 0.1097 & 0.0245  \\ 
    \bottomrule
\end{tabular}
\vspace{0pt}
\end{table}

\add{\subsection{Computational Cost vs. Performance}
Using an LLM backbone increases computational cost compared to lightweight models such as LSTM.
On a single NVIDIA RTX-3090, in our experiments, LSTM training completes (with early stopping) in approximately 30 minutes, whereas the LLM-based variants (TimeLLM and +SAP) require approximately two days under the same training protocol (early stopping with patience $=5$ and a maximum of $15$ epochs).
Averaged over the six TW21 stocks in Table~\ref{tab:genera_future1}, +SAP reduces MAE by 56.14\% and MSE by 73.23\% relative to LSTM.
Compared with the LLM backbone alone (TimeLLM), +SAP reduces MAE by 3.60\% and MSE by 6.40\% on average, with additional computational overhead relative to the base LLM.
For daily forecasting, the inference latency aligns with the prediction frequency; therefore, the cost--performance trade-off is determined by the application’s accuracy requirements and available compute budget.}
\section{Conclusion and Future Directions}
\label{sec:conclusion}

In this study, we introduced a multi-stock training framework designed to integrate news sentiment for stock price prediction while mitigating the impact of irrelevant information.
We implemented and evaluated three attention-based pooling mechanisms: self-attentive (+SAP), cross-attentive (+CAP), and position-aware self-attentive pooling.
The approach utilizes stock names to guide these mechanisms, enabling the model to filter information from real-world news feeds in the Taiwan stock market.
The empirical results on the TW21 dataset indicate that the framework reduces prediction errors, and its evaluation on the BigData23 dataset demonstrates its applicability across different data environments.

Our comparative analysis across 1-day and 5-day prediction horizons, using both LLaMA and GPT-2 as backbone models, yielded the following observations.
The +SAP mechanism exhibited lower error variance and consistently reduced prediction errors across stocks.
In contrast, while the +CAP mechanism yielded low errors for specific large-cap stocks, its metrics exhibited higher variance, indicating a sensitivity to hyperparameter tuning that warrants further investigation.
Furthermore, incorporating the ``Name Prompt'' did not yield a further reduction in error metrics, suggesting that the attention mechanisms alone are sufficient for identifying relevant news content within our framework.

Building on these findings, we outline directions for future research.
First, regarding data scope and generalization limits, we intend to expand the TW21 dataset to include a broader range of securities, specifically targeting small-cap and low-visibility stocks.
This will allow us to evaluate the model's applicability beyond large-cap technology companies and analyze the utility of semantic embeddings in less information-rich environments.
We will also explore the integration of longer-term historical data, such as 60- or 120-day input windows, to forecast medium-term price trends (e.g., the 20-day moving average).
This will assess the model's capability to capture both short-term volatility and long-term momentum.

Second, to address the limitations observed in ticker-based datasets (such as BigData23), we plan to investigate Entity Linking and knowledge retrieval techniques.
By mapping ambiguous tickers (e.g., ``F'') to their corresponding semantic entities (e.g., ``Ford Motor Company'') before feeding them into the LLM, we aim to provide the contextual representation required for news filtering.
Additionally, we will experiment with incorporating features from traditional econometric models, such as ARIMA, to augment the model's input representation.

Finally, future work will involve backtesting these predictive models as components of trading strategies to evaluate their performance against established financial benchmarks.

\section*{ACKNOWLEDGMENT}
We thank Ming-Chi Yen for maintaining the GPU infrastructure. Gratitude is extended to TAIDE (LLM) and twstock (stock data) for their resources. Gemini 3.1 Pro and GPT-5.2 Thinking were used for language refinement.

\bibliographystyle{JISEbib}
\bibliography{references}

\vspace{-4\baselineskip}

\begin{IEEEbiography}[{\includegraphics[width=3cm,height=4cm]{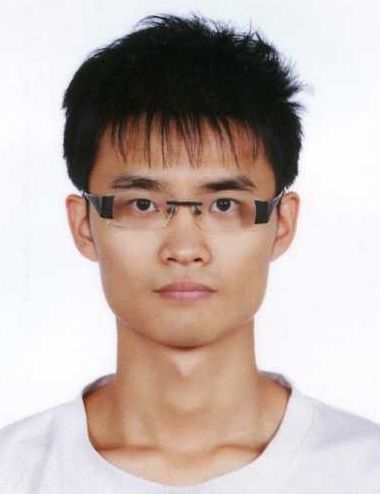}}]{Pei-Jun Liao}
received the B.S. degree from the Department of Electrical Engineering, National Taiwan University, Taipei, Taiwan, and the M.S. degree from the Graduate Institute of Communication Engineering, National Taiwan University. 
He is currently pursuing a Ph.D. degree at the Graduate Institute of Electrical Engineering, National Taiwan University. 
His research interests include speech processing and self-supervised learning.
\end{IEEEbiography}

\vspace{-4\baselineskip}

\begin{IEEEbiography}
[{\includegraphics[width=3cm,height=4cm]{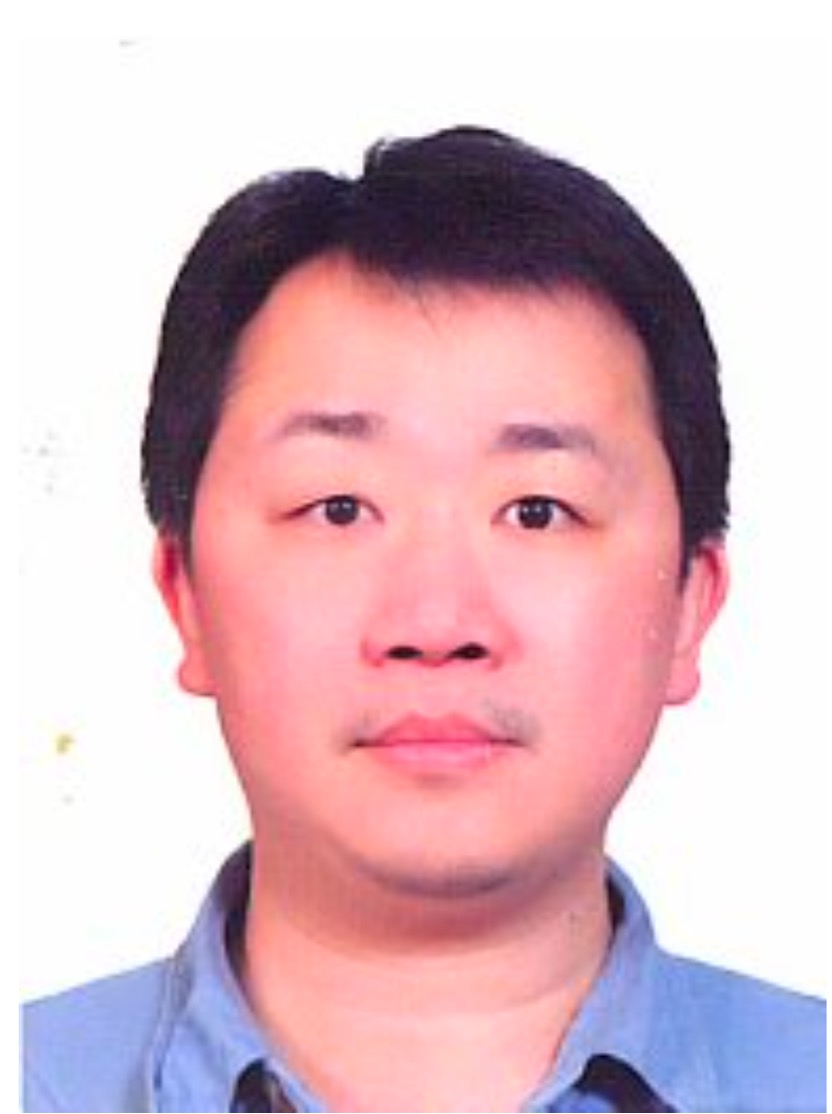}}]{Hung-Shin Lee} received the Ph.D. degree in electrical engineering from National Taiwan University (NTU), Taipei, Taiwan, in 2021. He is currently an AI Engineer with United Link Co., Ltd., Taiwan. Since February 2026, he has also been an Adjunct Assistant Professor with Graduate Institute of AI Interdisciplinary Applied Technology, National Taiwan Normal University, Taipei, Taiwan. His research interests include speech recognition and synthesis, as well as speech and video enhancement.
\end{IEEEbiography}

\vspace{-4\baselineskip}

\begin{IEEEbiography}[{\includegraphics[width=3cm,height=4cm]{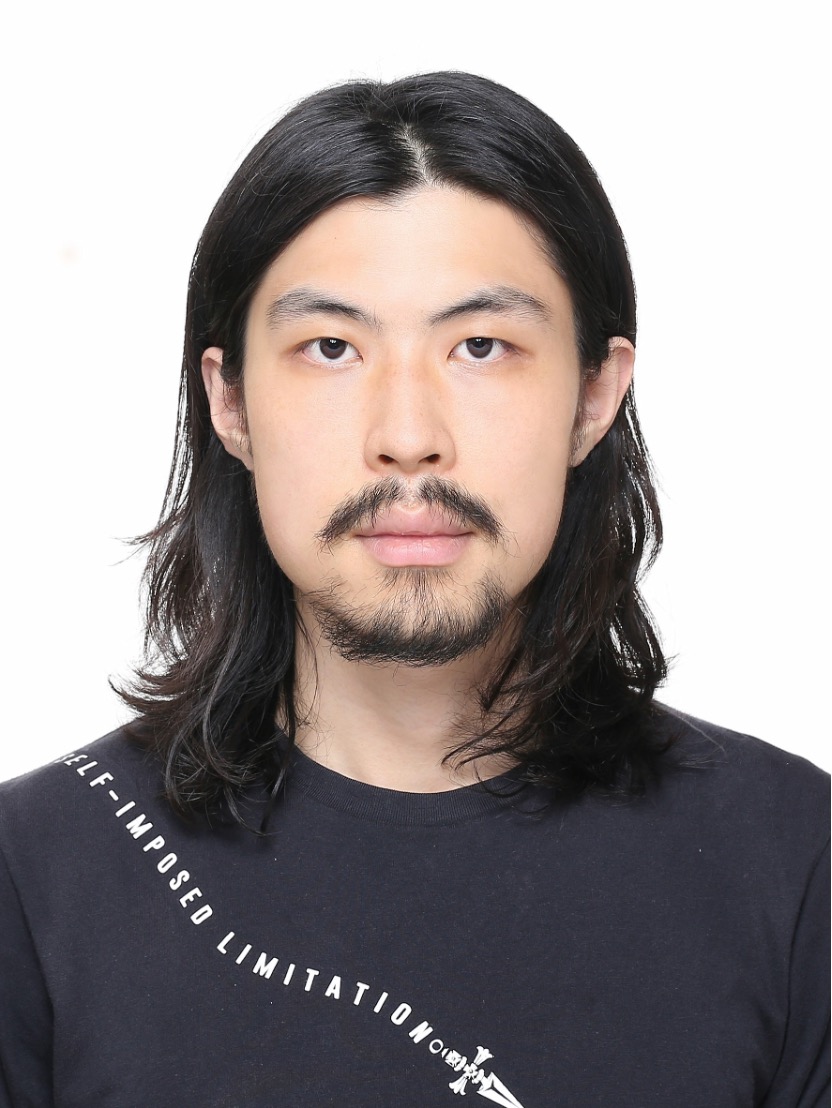}}]{Yao-Fei Cheng}
received the B.S. degree from the Department of Computer Science and Information Engineering, National Taipei University of Technology, Taipei, Taiwan, and the M.S. degree in Computational Linguistics from the University of Washington, Seattle.
His research interests include speech processing, natural language processing, and self-supervised learning.
\end{IEEEbiography}

\begin{IEEEbiography}[{\includegraphics[width=3cm,height=4cm]{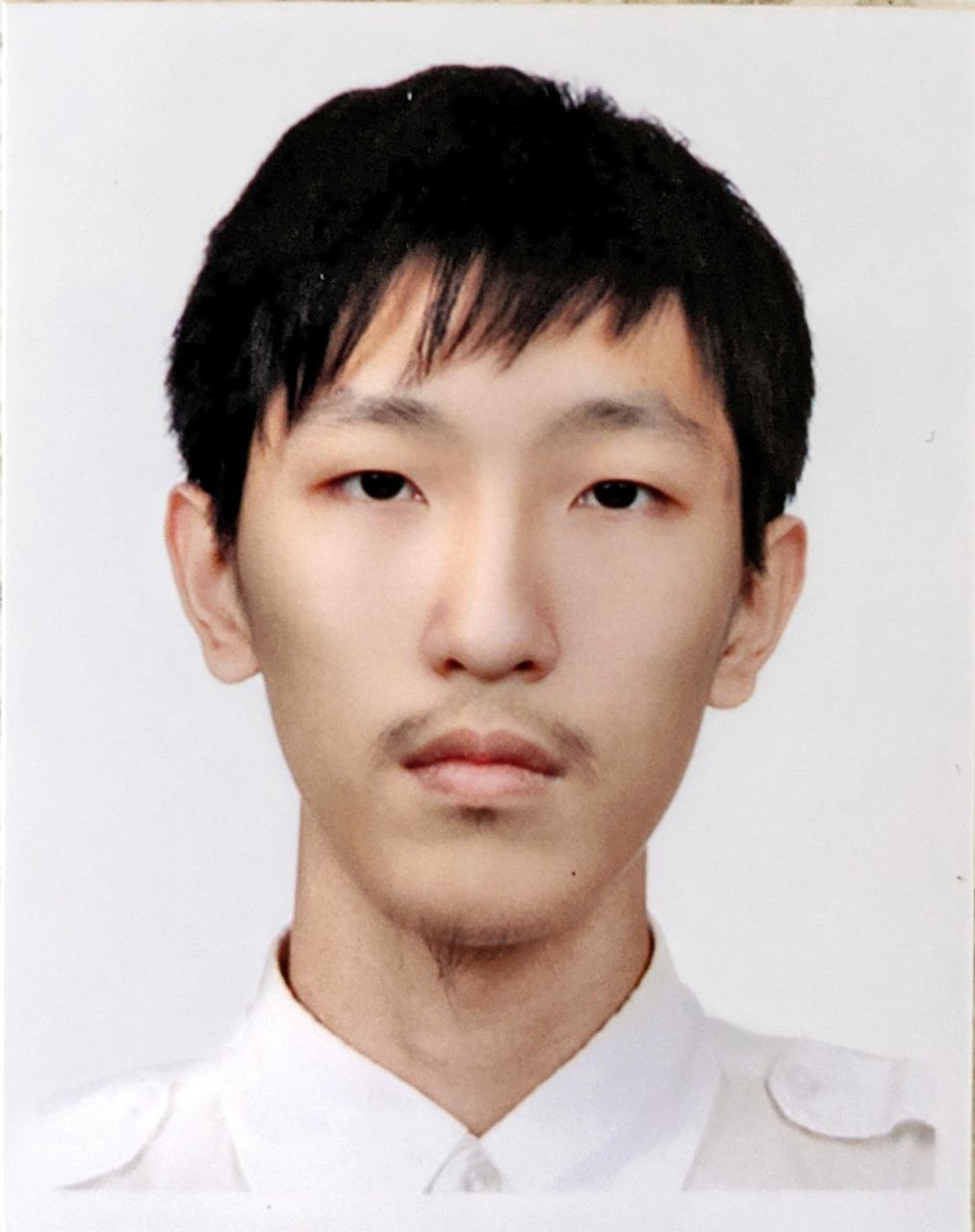}}]{Li-Wei Chen}
received the B.S. degree from the Department of Computer Science and Information Engineering, National Taipei University of Technology, Taipei, Taiwan. 
He is currently pursuing a M.S. degree at the Department of Computer Science, National Tsing Hua University, Hsinchu, Taiwan. 
His research interests include speech enhancement and speech synthesis.
\end{IEEEbiography}

\vspace{-4\baselineskip}

\begin{IEEEbiography}[{\includegraphics[width=3cm,height=4cm]{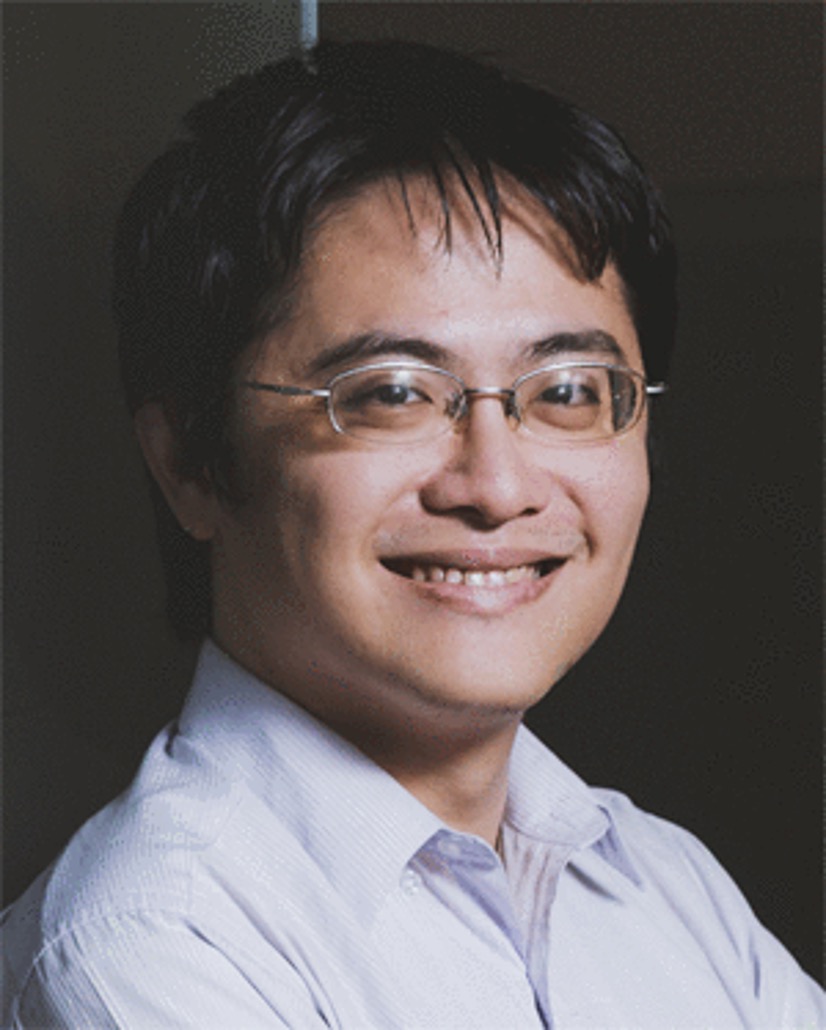}}]{Hung-yi Lee} is a professor of EE and CS at National Taiwan University (NTU). His recent research focuses on developing technology that can reduce the requirement of annotated data for speech processing and natural language processing. He is a co-organizer of the workshop "Self-Supervised Learning for Speech and Audio Processing" at NeurIPS (2020) and AAAI (2022), special issue "New Trends in self-supervised speech processing" at INTERSPEECH (2020), workshop "Meta Learning and Its Applications to Natural Language Processing" at ACL (2021). He was the technical program co-chair of ISCSLP 2022, technical program co-chair of INLG 2023 and program chair of Machine Learning Summer School (MLSS) 2021 Taipei. He was a lead guest editor of the Journal of Selected Topics in Signal Processing (JSTSP), editing a special issue, "Self-Supervised Learning for Speech and Audio Processing".
\end{IEEEbiography}

\vspace{-4\baselineskip}

\begin{IEEEbiography}[{\includegraphics[width=3cm,height=4cm]{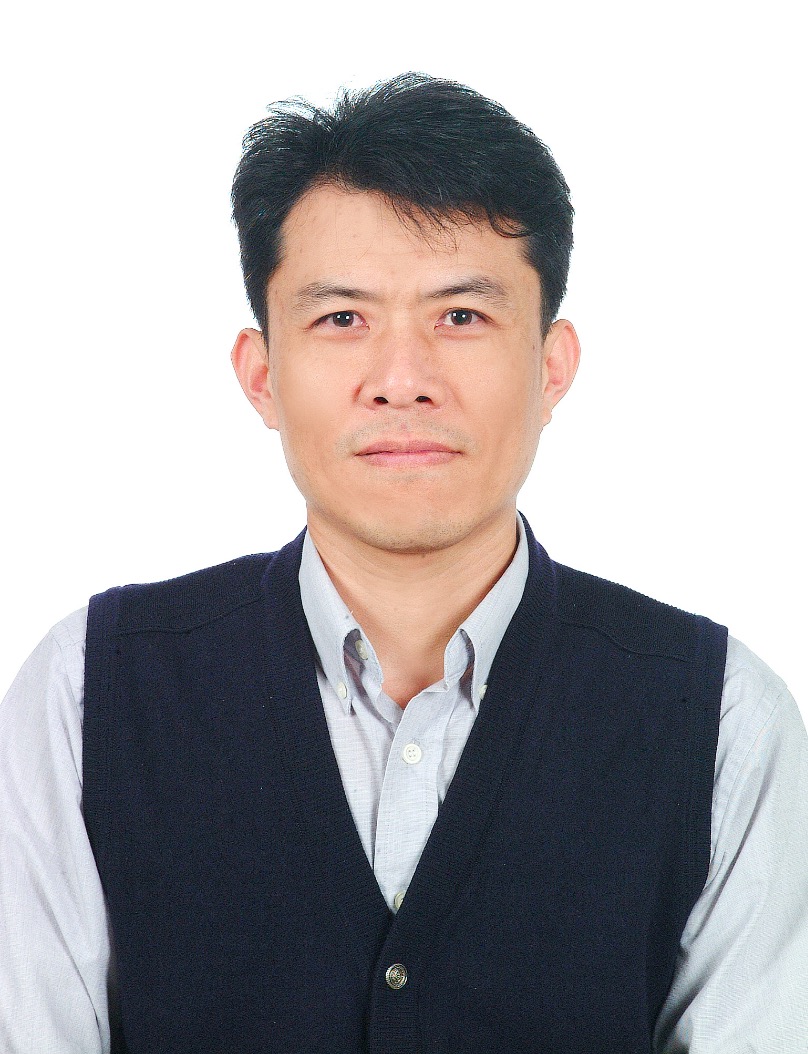}}]{Hsin-Min Wang} received the B.S. and Ph.D. degrees in electrical engineering from National Taiwan University, Taipei, Taiwan, in 1989 and 1995, respectively. In October 1995, he joined the Institute of Information Science, Academia Sinica, Taipei, Taiwan, where he is currently a Research Fellow. He was a Joint Professor in the Department of Computer Science and Information Engineering at National Cheng Kung University from 2014 to 2023. He was an Associate Editor of IEEE/ACM Transactions on Audio, Speech and Language Processing from 2016 to 2020. He currently serves a Senior Editor of APSIPA Transactions on Signal and Information Processing. His major research interests include spoken language processing, natural language processing, multimedia information retrieval, and machine learning. He was a General Co-Chair of ISCSLP2016, ISCSLP2018, ASRU2023, and O-COCOSDA2024 and a Technical Co-Chair of ISCSLP2010, O-COCOSDA2011, APSIPAASC2013, ISMIR2014, ASRU2019, and APSIPAASC2023. He received the Chinese Institute of Engineers Technical Paper Award in 1995 and the ACM Multimedia Grand Challenge First Prize in 2012. He was an APSIPA distinguished lecturer for 2014–2015. He is a member of ISCA, ACM, and APSIPA.
\end{IEEEbiography}

\end{document}